\documentclass[12pt]{article}
\usepackage[latin9]{inputenc}
\usepackage{geometry}
\usepackage{amsmath}
\usepackage{statmath}
\usepackage{mathtools}
\usepackage{amssymb}
\usepackage{stmaryrd}
\usepackage{etoc}
\usepackage{setspace}
\usepackage[authoryear]{natbib}
\usepackage{minitoc}
\usepackage[unicode=true,
 bookmarks=false,
 breaklinks=false,pdfborder={0 0 1},backref=section,colorlinks=false]
 {hyperref}
\hypersetup{
 colorlinks,linkcolor=red,citecolor=blue}

\makeatletter

\usepackage{amsfonts}
\usepackage{amsthm}
\usepackage{lscape}
\usepackage[toc,page,header]{appendix}
\usepackage{dcolumn}
\usepackage{rotating}
\usepackage{comment}
\usepackage{color}

\usepackage{diagbox}
\usepackage[flushleft]{threeparttable}
\usepackage{subfigure}
\usepackage{booktabs}
\usepackage{multirow}    
\usepackage{setspace}
\usepackage{bbm}
\usepackage{amsfonts} 
\usepackage{graphicx}

\usepackage{caption}
\usepackage{amssymb}
\usepackage{subcaption}
\usepackage{amsmath}
\usepackage{pifont}
\usepackage{etoc}

\usepackage{pgf}
\usepackage{tikz}\usepackage{caption}
\usepackage{tkz-tab}
\usepackage{tikz-3dplot}
\usetikzlibrary{calc}
\usetikzlibrary{arrows, automata}

\usepackage{changepage}
\usepackage{booktabs}
\usepackage{graphicx}

\usepackage{hyperref}

\newcolumntype{d}[1]{D{.}{.}{#1}}
\newcolumntype{t}[1]{D{,}{,}{#1}}
\newcolumntype{i}[1]{D{.}{}{#1}}

\newtheorem{assumption}{Assumption}

\newtheorem{proposition}{Proposition}

\theoremstyle{plain}

\numberwithin{equation}{section}

\usepackage{xcolor}

\newcommand{\Expectation}{\mathbb{E}}
\newcommand{\Cov}{\mathbb{C}\mathrm{ov}}
\newcommand{\AsyCov}{\operatorname{AsyCov}}

\newcommand{\Prob}{\mathbb{P}}

\makeatother

\begin{document}
\doparttoc % Tell to minitoc to generate a toc for the parts
\faketableofcontents % Run a fake tableofcontents command for the partocs
\title{Inference for the Marginal Value of Public Funds\thanks{Vohra (\href{mailto:vohra@ucsd.edu}{vohra@ucsd.edu}): UC San Diego. I am grateful to Julie Cullen, Graham Elliott, Itzik Fadlon, Stefan Faridani, Jacob Goldin, Xinwei Ma, Craig McIntosh, Harrison Mitchell, and Kaspar Wuthrich for helpful comments. An earlier draft of this paper was titled ``Valid Inference on Functions of Causal Effects in the Absence of Microdata.''}}
\author{Vedant Vohra}
\date{\today}
\maketitle
\vspace{-0.2cm}
\thispagestyle{empty}

\begin{abstract}
Economists often estimate causal effects of policies on multiple outcomes and summarize them into scalar measures of cost-effectiveness or welfare, such as the Marginal Value of Public Funds (MVPF). In many settings, microdata underlying these estimates are unavailable, leaving researchers with only published estimates and their standard errors. We develop tools for valid inference on functions of causal effects, such as the MVPF, when the correlation structure is unknown. Our approach is to construct worst-case confidence intervals, leveraging experimental designs to tighten them, and to assess robustness using breakdown analyses. We illustrate our method with MVPFs for eight policies. (\textit{JEL} C12, C21, H00)
\end{abstract}

\newpage 
\setcounter{page}{1}
\section{Introduction\label{sec:Introduction}}

Policymakers increasingly rely on evidence from randomized evaluations to guide decisions about which programs to fund and scale. These evaluations often report multiple estimated causal effects---for example, a tax credit program's impacts on earnings, after-tax income, and labor force participation---but policymakers typically care about a summary measure that aggregates these effects into a measure of overall cost-effectiveness of the policy, such as the Marginal Value of Public Funds (MVPF). Determining whether a program delivers ``bang for the buck'' thus requires inference on a scalar function of several estimated causal effects, rather than on any individual effect alone.

Conducting valid inference on such summary measures of policy cost-effectiveness is often difficult in practice. In many cases, researchers observe only the estimated causal effects and their standard errors but lack information about the correlations between them. These correlations are crucial for quantifying uncertainty: the variance of any scalar function that aggregates multiple effects depends not only on the precision of each individual estimate but also on how the estimates co-vary. If the underlying microdata were available, the covariances between causal effect estimates could be computed directly, allowing the variance of the function to be estimated \citep[e.g.,][]{zellner1962efficient}. However, in many settings of practical interest, the microdata are unavailable for ex-post analysis, leaving researchers to rely only on published estimates and their standard errors. In this paper, we study the problem of inference on functions of multiple causal effects when the correlation structure across these causal effects is unknown.

To illustrate the challenge, we focus on the problem of conducting inference for the Marginal Value of Public Funds (MVPF) of a policy \citep{hendren2020unified}. The MVPF is a widely used metric for evaluating the welfare consequences of government expenditure. It is defined as a non-linear function of multiple causal effects: the benefits a policy provides to its recipients are divided by the policy's net cost to the government. \citet{hendren2020unified} construct MVPFs for more than one hundred policies using causal effects reported in existing studies. In most cases, only the point estimates and their standard errors are available to them, while the microdata underlying these estimates are inaccessible for such ex-post analysis. The challenge for inference is that the variance of the MVPF depends on the correlations across causal effects, which are not reported and not estimable.

We propose a simple inference procedure that delivers valid confidence intervals for functions of causal effects, even when the correlation structure across effects is unknown. The idea is straightforward: we ask what is the largest possible variance of the function given the available information and we identify the correlation structure under which this upper bound is attained. Using this worst-case variance, we construct conservative confidence intervals that guarantee valid coverage. We show that this conservative approach enables meaningful inference when computing the confidence intervals for MVPF estimates. Importantly, we formulate the problem of finding the variance upper bound as an optimization problem, which also makes it straightforward to incorporate other setting-specific information---for example, known independence between causal effects---to further tighten the confidence intervals and improve statistical precision.

Second, we show how confidence intervals can be tightened further still when the causal effects correspond to the impacts of a randomized treatment on multiple outcomes. In this setting, we characterize the off-diagonal entries of the covariance matrix and show that they take a particularly interpretable form, the sign of which may be known from prior studies, economic theory, or other data sources. Incorporating such information allows us to meaningfully increase statistical power to reject null hypotheses of interest. 

Finally, we introduce a complementary approach to inference and ask a different question from worst-case inference. Instead of focusing on the largest possible variance given the available information, we ask how sensitive a policy-relevant conclusion is to uncertainty about the correlation structure. For example, a policymaker may wish to test whether a dollar spent on a policy provides beneficiaries with at least one dollar of benefits, i.e., $H_0: \text{MVPF} < 1$ against $H_1: \text{MVPF} \geq 1$. We introduce a ``breakdown statistic'' that quantifies how robust this conclusion is to different correlation structures: it measures the proportion of admissible correlation structures under which the null hypothesis would \emph{not} be rejected. The statistic takes values between 0 and 1, where a value of 0 implies that we can conclude that the MVPF is greater than 1 under all plausible correlation structures, while a value of 1 implies that we cannot reject the null under any correlation structure. Unlike inference based on the worst-case variance, which guarantees valid coverage but might be conservative, the breakdown statistic facilitates comparisons of the robustness with which we can arrive at a policy conclusion---for example, whether the MVPF of a policy is greater than 1---across settings.

We illustrate our inference procedure by conducting inference on the MVPF for eight different policies. First, we show that meaningful inference is possible even in the absence of \emph{any} microdata, using the upper bound of the variance alone. Second, \citet{hendren2020unified} note that because the MVPF reflects the shadow price of redistribution, a welfare-maximizing government should have a positive willingness-to-pay to reduce the statistical uncertainty around the cost of redistribution. We demonstrate how this uncertainty can be reduced by leveraging setting-specific information about the sign of correlations across outcomes. In fact, our novel characterization of the covariance structure in randomized trials allows us to tighten MVPF confidence intervals beyond the worst-case by up to 30\% in the policies we consider. Finally, we compute the breakdown statistic for the MVPF across multiple policies and illustrate how this metric can guide policymakers choosing among alternative policies.

Our work contributes to the literature on welfare analyses of government expenditure \citep[e.g.,][]{chetty2009sufficient,heckman2010rate,hendren2020unified}. While existing tools provide a unified framework for evaluating the welfare consequences of government policies, statistical methods for conducting inference on welfare metrics under frequently encountered data limitations have been less developed. Our inference procedures strengthen the MVPF framework by providing a formal approach to quantifying statistical uncertainty in welfare metrics. \citet{hendren2020unified} show that increasing spending on Policy A is welfare-improving by reducing spending on Policy B if and only if the MVPF of Policy A exceeds that of Policy B; the methods developed in this paper provide a valid test for the policy-relevant null hypothesis, $H_0: \text{MVPF}_A < \text{MVPF}_B$.

\citet{hendren2020unified} propose a parametric bootstrap approach that constructs confidence intervals for the MVPF under a user-specified correlation structure. While such an approach can yield valid inference when the specified structure is indeed the worst case, misspecification may lead to confidence intervals that fail to achieve nominal coverage. Our method avoids this risk by formally identifying---rather than assuming---the correlation structure that maximizes the variance, solving an optimization problem that guarantees valid inference regardless of the true correlation structure. Moreover, our setup allows us to incorporate additional setting-specific information---for example, known independence across estimates or theory-driven sign restrictions---thereby increasing statistical power when such information is available.

Our methods might be applicable beyond the MVPF as well, in other settings when the correlation structure across causal effects might be difficult to obtain. First, researchers are frequently interested in functions of causal effects reported in existing publications, but the underlying microdata may be inaccessible. This can occur when the effects are estimated using privately held administrative data or when replication files are not publicly released. Replication data are missing for nearly half of all empirical papers published in the \emph{American Economic Review} \citep{christensen2018transparency}, underscoring how common this problem is. Second, even when the underlying data are technically available, computing the correlations can be prohibitively costly when the effects come from distinct datasets with common units but difficult-to-merge identifiers. For example, unique identifiers may be missing in historical decennial Census data \citep{ruggles2018historical}, or the relevant data sources may be stored across separate federal agencies, as is the case when linking administrative tax data and administrative crime records for the full population \citep{rose2018effects}.

Finally, \citet{cocci2023standard} develop a closely related procedure in a different context: bounding the asymptotic variance of an estimate for a structural parameter when calibrating models to match empirical moments. Their work shows how to compute worst-case standard errors when the off-diagonal elements of the variance-covariance matrix are unknown, using only the variances of the empirical moments. We extend their framework in three important ways. First, we exploit the structure of randomized treatments to characterize the covariance matrix, which enables us to impose theory-motivated sign restrictions and potentially yields substantial power improvements. Second, our paper applies this variance-bounding approach to a new domain---inference on welfare metrics, such as the MVPF---where analysts frequently lack access to the underlying microdata. Third, we introduce a ``breakdown'' approach that quantifies the robustness of policy conclusions to uncertainty about the correlation structure.

\section{Setting}\label{section:setting}

Our starting point is a vector of estimated causal effects, denoted by $\bm{\hat{\beta}} \in \mathbb{R}^d$, that we seek to aggregate into a measure of the cost-effectiveness of a policy. We assume that $\bm{\hat{\beta}}
$ asymptotically follows a joint Normal distribution with variance-covariance matrix $\bfV$. Since it is standard practice to report standard errors for individual estimates, we assume we have access to consistent estimates of the diagonal entries of $\bfV$. In contrast, the covariances between estimated causal effects---the off-diagonal entries of $\bfV$---are rarely reported. We focus on a setting where the underlying microdata are unavailable, so these off-diagonal entries cannot be directly estimated. We summarize the available information in Assumption \ref{assu:maintained-assumption}. 

\begin{assumption}\label{assu:maintained-assumption}
From an existing study, we observe a vector of estimated causal effects 
$\bm{\hat{\beta}} = \big(\widehat{\beta}_1,\dots,\widehat{\beta}_d\big)$ 
for the true causal effects 
$\bm{\beta} = \big({\beta}_1,\dots,{\beta}_d\big)$, 
along with their corresponding standard errors 
$\bm{\hat{\sigma}} = \big(\widehat{\sigma}_1,\dots,\widehat{\sigma}_d\big)$.
We assume:
\begin{enumerate}
    \item[(i)] Consistency: $\bm{\hat{\beta}} \xrightarrow{p} \bm{\beta}$.
    \item[(ii)] Asymptotic normality: 
    $\sqrt{n}\big(\bm{\hat{\beta}} -  \bm{\beta}\big) 
    \xrightarrow{d} \mathcal{N}\big(0,\mathbf{V}\big)$, 
    where $\mathbf{V}$ is a $d \times d$ positive semi-definite variance-covariance matrix.
    \item[(iii)] Consistent standard errors: 
    $\bm{\hat{\sigma}^2}$ consistently estimates the diagonal entries of $\mathbf{V}$, i.e. 
    $\big(\widehat{\sigma}_1^2,\dots,\widehat{\sigma}_d^2\big) 
    \xrightarrow{p} \big(\sigma_1^2,\dots,\sigma_d^2\big)$.
\end{enumerate}
\end{assumption}

In our setting, $f(\bm{\hat{\beta}})$ represents an estimate of a policy's cost-effectiveness. We assume that the function aggregating the causal effects, $f: \mathbb{R}^d \to \mathbb{R}$, is continuously differentiable at $\bm\beta$, with  $f'(\bm\beta)\neq0$. We place no additional restrictions on $f(\cdot)$ and allow it to be non-linear, since in practice its form depends on the economic mapping between the estimated causal effects and the cost-effectiveness measure for the policy being analyzed. This is summarized in Assumption \ref{assu:fn}. 
\begin{assumption}\label{assu:fn}
The function $f: \mathbb{R}^d \to \mathbb{R}$ is continuously differentiable at $\bm\beta$ and satisfies $f'(\bm\beta)\neq0$.
\end{assumption}

In summary, Assumption \ref{assu:maintained-assumption} states that we observe consistent estimates of the causal effects and their standard errors, but lack reliable information about the covariances between them. Assumption \ref{assu:fn} requires that the function mapping estimated causal effects into the cost-effectiveness measure is smooth and well-behaved. We maintain Assumptions \ref{assu:maintained-assumption} and \ref{assu:fn} throughout the paper. 

Under these assumptions, we apply the delta method to obtain the asymptotic distribution of $f(\bm{\hat{\beta}})$:
\[
    \sqrt{n}\Big(f(\bm{\hat{\beta}}) - f(\bm{\beta})\Big) \xrightarrow{d} \mathcal{N}(0,\tau^2)
\]
where
\[
    \tau^2 = \sum_{i=1}^{d}\sum_{j=1}^{d} \sigma_{ij} 
    \frac{\partial f(\bm{\beta})}{\partial \beta_i} 
    \frac{\partial f(\bm{\beta})}{\partial \beta_j}
    = \left[ \sum_{i=1}^{d}\left(\sigma_i \frac{\partial f(\bm{\beta})}{\partial \beta_i}\right)^2 \right]
    + \Bigg[\mathop{\sum\limits_{i=1}^{d}\sum\limits_{j=1}^{d}}_{\{i,j : i \neq j\}} 
    \sigma_{ij} \frac{\partial f(\bm{\beta})}{\partial \beta_i} 
    \frac{\partial f(\bm{\beta})}{\partial \beta_j}\Bigg]
    \tag{2.1}\label{eqn:target-var}
\]
and $\sigma_{ij}$ denotes the covariance between $\beta_i$ and $\beta_j$.\footnote{
The delta method relies on a first-order (linear) approximation of the function $f(\cdot)$ around $\bm{\beta}$. Under the maintained assumptions, this approximation is valid asymptotically. However, in finite samples, if the variance of $\bm{\hat{\beta}}$ is large, $\bm{\hat{\beta}}$ may deviate from $\bm{\beta}$ with non-negligible probability, making the linear approximation less accurate.} We define
\[
    \rho_{ij} \equiv \frac{\sigma_{ij}}{\sigma_i \sigma_j},
\]
as the correlation coefficient between $\beta_i$ and $\beta_j$.

The objective of this paper is to learn about $\tau^2$, the asymptotic variance of $f(\bm{\hat{\beta}})$. The central challenge is that $\tau^2$ depends on the covariances $\sigma_{ij}$, which are not estimable in our setting because the underlying microdata are unavailable. This raises the key question: what can be learned about $\tau^2$ when $\sigma_{ij}$ for $i \neq j$ cannot be estimated?

\section{Inference Procedure}\label{section:inference-procedure}

Since the asymptotic variance of $f(\bm{\hat{\beta}})$ depends on the correlation structure across the estimated causal effects---and this correlation structure cannot be estimated in the absence of microdata---we consider an alternative approach to inference on $f(\bm\beta)$. We ask: given the observed information, how large could the asymptotic variance of $f(\bm{\hat{\beta}})$ be? We then use an estimate of this variance upper bound to conduct valid hypothesis tests on $f(\bm\beta)$.

To motivate this approach and provide a rationale for focusing on the variance upper-bound, consider testing:
\begin{align}\label{eqn:hypothesis}
    H_0: f(\bm\beta) \leq k 
    \quad \text{against} \quad 
    H_1: f(\bm\beta) > k.
\end{align}
When the variance $\tau^2$ can be consistently estimated, standard $t$-tests control size and are uniformly most powerful. The difficulty arises when the correlations across effects ($\rho_{ij}$) are unknown and $\tau^2$ cannot be estimated. In this case, the problem can be framed as hypothesis testing with nuisance parameters,  $\rho_{ij}$ for $i\neq j$. Finding the UMP test in this setting corresponds to identifying the least favorable
distribution of the nuisance parameters \citep[Theorem 3.8.1 in][]{romano2005testing}.\footnote{A least favorable distribution is the distribution on the nuisance parameters under which the test performs the worst, or in other words, the distribution under which the probability of correctly rejecting a false null is smallest. If a test controls size and has good power even under this ``worst-case'' scenario, then it will perform at least as well under all other admissible distributions.} While least
favorable distributions are often challenging to characterize \citep[see, e.g.,][]{elliott2015nearly},  our setting is simplified by the fact that the nuisance parameters enter the distribution of $f(\hat{\bm\beta})$ only through its variance. Since power is minimized when variance is maximized, finding the least favorable distribution---and hence the uniformly most powerful test---corresponds to finding the correlation structure that maximizes $\tau^2$. This provides the statistical rationale for focusing on the variance upper bound.

In Section~\ref{section:general-case-inference}, we consider the general setting where no additional structure is imposed on the correlation structure, so the variance upper bound is determined solely by the mathematical constraints of the correlation matrix. In Section~\ref{section:sign-constraints}, we specialize to cases where treatment assignment is either completely randomized or randomized conditional on observed covariates. In this setting, we provide a novel characterization of the covariance structure that allows us to incorporate information from prior studies, economic theory, or other data sources to impose sign constraints on elements of the covariance matrix.

\subsection{Worst-Case Inference}\label{section:general-case-inference}

We can re-express Equation~\eqref{eqn:target-var} in terms of the correlations $\rho_{ij}$ as follows:
\begin{align}\label{eqn:target-var-corr}
    \tau^2 
    &= \sum_{i=1}^{d}\left(\sigma_i \frac{\partial f(\bm{\beta})}{\partial \beta_i}\right)^2 
    + \mathop{\sum_{i=1}^{d}\sum_{j=1}^{d}}_{\{i,j : i \neq j\}}
    \rho_{ij}\,\sigma_i \sigma_j 
    \frac{\partial f(\bm{\beta})}{\partial \beta_i} 
    \frac{\partial f(\bm{\beta})}{\partial \beta_j}.
\end{align}
Since $\beta_i$ and $\sigma_i$ can be consistently estimated from the observed data, obtaining an upper bound for $\tau^2$ amounts to maximizing Equation~\eqref{eqn:target-var-corr} with respect to $\rho_{ij}$ for $i \neq j$, subject to certain constraints. We formulate it as the following convex optimization problem:
\begin{align*} 
\label{worst-case-sdp} \tag*{\textbf{SDP.1}} 
\underset{ \{\rho_{ij}\}_{i,j=1}^{d} }{\text{Maximize}} \qquad & \tau^2  \\
\text{subject to}  \qquad & \mathbf{V} \succeq 0 \tag{C.1}  \label{eqn:psd-const} \\
& \rho_{ij} = \rho_{ji}   &  \forall \ i,j=1,\dots,d \tag{C.2} \label{eqn:symmetry-const} \\
& \rho_{ij} \in [-1,1]    &  \forall \ i,j=1,\dots,d \tag{C.3} \label{eqn:corr-range-const} \\
& \rho_{ii} = 1    &  \forall \ i=1,\dots,d \tag{C.4} \label{eqn:corr-diag-const}
\end{align*}

Constraint~\eqref{eqn:psd-const} requires the variance-covariance matrix to be positive semidefinite; Constraint~\eqref{eqn:symmetry-const} enforces symmetry of the variance-covariance matrix; and Constraint~\eqref{eqn:corr-range-const} ensures that all pairwise correlations lie within $[-1,1]$. Problem~\eqref{worst-case-sdp} is therefore a well-defined semidefinite program (SDP) that can be solved using existing optimization tools \citep{grant2008graph,grant2014cvx}. A key advantage of this formulation is its flexibility: we can incorporate available information about the correlations as additional constraints. For instance, in some cases, it may be known that two estimates are uncorrelated, such as when they are constructed using independent, non-overlapping samples. This information can be incorporated into the 
optimization problem by fixing the corresponding correlation to be zero. This flexibility is particularly important for the analysis in Section~\ref{section:sign-constraints}, where we leverage our characterization of the off-diagonal entries of the covariance matrix to impose theory-motivated sign restrictions.

We denote the maximum variance obtained by solving \ref{worst-case-sdp} as $\tau^2_{\max}$ and note the following. First, confidence intervals constructed using $\tau_{\max}$ will, by construction, have weakly higher coverage probability than those based on $\tau$. While this reduces power, it guarantees size control; coverage is exact only when $\tau_{\max} = \tau$. Second, in settings where estimating the covariance across estimates is feasible but costly, we recommend that researchers first test their hypotheses using $\tau_{\max}$. Rejecting a null hypothesis under the worst-case variance implies that the null would also be rejected using the true variance. This allows researchers to conduct valid inference while avoiding the costs of computing the full covariance matrix. Finally, in Appendix Section~\ref{section:mikkel-inference}, we compare our approach to that of \citet{cocci2023standard} who propose a method for worst-case inference when matching structural parameters to empirical moments in overidentified settings. We show how tighter bounds can be found than what is implied by Lemma 1 in \cite{cocci2023standard}, and illustrate through an example why maximizing the variance is challenging even in the simple case where there are no additional constraints beyond those in \ref{worst-case-sdp}. 

\subsection{Worst-Case Inference Under Random Treatment Assignment}\label{section:sign-constraints}

In contrast to the generic case considered in Section~\ref{section:general-case-inference}, this section leverages information about treatment assignment to derive more powerful tests on cost-effectiveness parameters. When treatment is randomized---either completely or conditional on observables---we obtain a novel, interpretable characterization of the covariance matrix. This characterization allows researchers to impose sign or independence restrictions on the correlation matrix, grounded in theory, prior evidence, or auxiliary data. Incorporating these restrictions can substantially sharpen variance bounds and deliver more precise inference on policy-relevant parameters.

Let $Y_{ij}(1)$ denote the treated potential outcome $j$ for unit $i$ and $Y_{ij}(0)$ denote the control potential outcome $j$ for unit $i$, where $j \in \{1,...,d\}$. Let $Z_i \in \{0,1\}$ indicate treatment assignment, where $Z_i = 1$ if unit $i$ is treated and $Z_i =0$ otherwise. We assume random assignment, meaning that treatment is independent of the full vector of potential outcomes:
\begin{align*}
 \Big(Y_{ij}(1),Y_{ij}(0)\Big) \perp Z_i   \quad \text{for all} \quad j=1,\dots,d
\end{align*}
The observed outcome is $Y_{ij} = Z_i\cdot Y_{ij}(1)+(1-Z_i)\cdot Y_{ij}(0)$. For each outcome $j$, the average treatment effect (ATE) is given by $\beta_j = \Expectation[Y_{ij}(1) - Y_{ij}(0)]$, and we estimate it using the difference in sample means between the treated and control groups:
\begin{align*}
    \widehat{\beta}_j = \frac{1}{n_1} \sum_{i: Z_i = 1} Y_{i,j} - \frac{1}{n_0} \sum_{i: Z_i = 0} Y_{i,j}
\end{align*}
where $n_1$ and $n_0$ are the number of treated and control units, respectively. Let $\hat{\bm\beta} \in \mathbb{R}^d$ denote the vector of estimated treatment effects. In this setting, the asymptotic variance-covariance matrix $\bfV$ has a structure that allows for a simple and interpretable characterization, summarized in the following proposition.

\begin{proposition}\label{prop:cov-result}
Let $\beta_p$ and $\beta_q$ denote the average treatment effects of a randomized treatment $Z_i \in \{0,1\}$ on two outcomes $Y_{ip}$ and $Y_{iq}$ respectively. Let the vector $\left\{\left(Y_{i p}(0), Y_{i p}(1), Y_{i q}(0), Y_{i q}(1), Z_i\right)\right\}_{i=1}^n$ be i.i.d. across units. Then, the asymptotic covariance, denoted by $\AsyCov(\cdot)$, between the difference-in-means estimators $\widehat{\beta}_p$ and $\widehat{\beta}_q$ is given by:
\begin{align*}
\AsyCov\Big(\widehat{\beta}_p, \widehat{\beta}_q\Big)
= \frac{\Cov(Y_{ip}, Y_{iq} \mid Z_i = 1)}{\Prob(Z_i = 1)}
+ \frac{\Cov(Y_{ip}, Y_{iq} \mid Z_i = 0)}{\Prob(Z_i = 0)}.
\end{align*}
\end{proposition}

The proof of Proposition~\ref{prop:cov-result} is in Appendix Section~\ref{appendix:cov-proof}. The proposition shows that the asymptotic covariance between estimated treatment effects $\widehat{\beta}_{p}$ and $\widehat{\beta}_{q}$ depends only on the covariances of outcomes $Y_{ip}$ and $Y_{iq}$ within the treatment and control groups. In particular, if the outcomes are positively correlated within both groups, the treatment effects on those outcomes must also move in the same direction. For example, in the case of a randomized tax credit expansion called Paycheck Plus, the MVPF depends on the effects of the program on after-tax income, earnings, and labor force participation.\footnote{The Paycheck Plus program is studied in \cite{miller2017expanding} and the MVPF for this program is computed in \cite{hendren2020unified}.} Since individuals with higher earnings also tend to have weakly higher after-tax income and are weakly more likely to participate in the labor force, Proposition~\ref{prop:cov-result} implies that the off-diagonal entries of $\bfV$ are non-negative. This information can be incorporated as additional constraints in the optimization problem \ref{worst-case-sdp}, thereby producing a (weakly) tighter upper bound. If it is known that all covariances between outcome pairs are non-negative as in the Paycheck Plus MVPF, we can solve the following optimization problem to obtain the variance upper bound:
\begin{align*} 
\label{sign-const-sdp} \tag*{\textbf{SDP.2}} 
\underset{ \{\rho_{ij}\}_{i,j=1}^{d} }{\text{Maximize}} \qquad & \tau^2  \\
\text{subject to}  \qquad & \mathbf{V} \succeq 0 \tag{C.1}  \label{eqn:psd-const} \\
& \rho_{ij} = \rho_{ji}  &  \forall \ i,j= 1 ,..., d \tag{C.2} \label{eqn:symmetry-const} \\
& \rho_{ij} \in [-1,1]  &  \forall \ i,j= 1 ,..., d\tag{C.3} \label{eqn:corr-range-const} \\
& \rho_{ii} = 1    &  \forall \ i= 1 ,..., d \tag{C.4} \label{eqn:corr-diag-const} \\
& \rho_{ij} \geq 0     &  \forall \ i,j= 1 ,..., d\tag{C.5} \label{eqn:sign-const}
\end{align*}
In Section \ref{section:application}, we show that adding Constraint \ref{eqn:sign-const} to the optimization problem reduces the width of the Paycheck Plus MVPF confidence intervals by nearly 30\%. 

We also extend Proposition \ref{prop:cov-result} to settings where treatment assignment is random only conditional on covariates, such that
$$
\left(Y_{i j}(1), Y_{i j}(0)\right) \perp Z_i \mid \mathbf{X}_i \quad \text{for all} \quad j = 1,\ldots,d
$$
where $\mathbf{X}_i \in \mathbb{R}^k$ is a vector of observed covariates. A similar characterization of the asymptotic covariance under unconfoundedness is provided in Appendix Section \ref{appendix:unconfoundedness}.

% In some settings, $\beta_i$ and $\beta_j$ will identify the ATE of a treatment on outcomes $Y_i$ and $Y_j$ with bounded support. When the outcome has bounded support, the support of the ATE will also bounded. In this case, one might be able to tighten the bounds on the variance further by incorporating this information into the optimization problem \ref{sign-const-sdp}. \cite{hossjer2022sharp} derive the following bounds for the covariance of two bounded random variables.

% \begin{remark}\label{remark:bounded-outcomes}
% Suppose that outcome $Y_i \in \Big[\underline{Y_i},\overline{Y_i}\Big]$ and $Y_j \in \Big[\underline{Y_j},\overline{Y_j}\Big]$. Define $\underline{\beta_i} = \underline{Y_i}-\overline{Y_i}$ and $\overline{\beta_i} = \overline{Y_i}-\underline{Y_i}$. Define $\underline{\beta_j}$ and $\overline{\beta_j}$ analogously. Then, the ATEs $\beta_i$ and $\beta_j$ will be bounded: $\beta_i\in \Big[\underline{\beta_i},\overline{\beta_i}\Big]$ and $\beta_j\in \Big[\underline{\beta_j},\overline{\beta_j}\Big]$. Then, the covariance of $\beta_i$ and $\beta_j$, $\sigma_{ij}$ satisfies:
% \begin{align*}
% & -\min\left[\Big(\beta_i-\underline{\beta_i}\Big)\Big(\beta_j-\underline{\beta_j}\Big),\Big(\overline{\beta_i}-\beta_i\Big)\Big(\overline{\beta_j}-\beta_j\Big)\right] \\
% & \leq \sigma_{ij} \\
% & \leq \min \left[\Big(\beta_i-\underline{\beta_i}\Big)\Big(\overline{\beta_j}-\beta_j\Big),\Big(\overline{\beta_i}-\beta_i\Big)\Big(\beta_j-\underline{\beta_j}\Big)\right] .
% \end{align*}    
% \end{remark}

\section{Breakdown Analysis}\label{section:breakdown}

In Section~\ref{section:inference-procedure}, we proposed a method that constructs worst-case confidence intervals for cost-effectiveness parameters, guaranteeing appropriate coverage regardless of the true correlation structure across causal effects. The strength of this method is its robustness: it delivers valid inference without making parametric assumptions about which correlation structures are more likely than others. The drawback is that tests based on the worst-case variance may leave policymakers underpowered to reject relevant null hypotheses. In this section, we develop a complementary approach to inference: we specify a set of plausible correlation structures, place a probability distribution over them, and ask: {how likely is it that a given hypothesis would be rejected if the true correlation structure were drawn from this set?} 

We begin by specifying three elements. First, we fix the null hypothesis of interest. For example, in the case of the MVPF, a policymaker may wish to test whether one dollar of government spending generates more than one dollar of benefits for recipients, i.e., $H_0:\,\text{MVPF} < 1$ vs. $H_1:\,\text{MVPF} \geq 1$. Second, we specify the set of admissible correlation structures. For instance, Proposition~\ref{prop:cov-result} may imply that correlations across the estimated causal effects are non-negative, so the admissible set is all correlation matrices with non-negative off-diagonal elements. Finally, we specify a probability distribution over this admissible set. For example, a policymaker might assume that all correlation structures in the admissible set are equally plausible. Alternatively, they might want to assume that correlation structures closer to independence are more plausible in their setting. We operationalize this by placing an LKJ prior \citep{lewandowski2009generating} over the admissible set. The LKJ distribution has density
    \(
    \pi(\rho) \propto \det(\rho)^{\eta - 1},
    \) 
where $\eta$ is the parameter governing which correlation structures are more likely than others. When $\eta = 1$, the prior is uniform over all correlation matrices. Larger values of $\eta$ place more mass near the identity matrix, favoring weaker correlations.

Next, we repeatedly draw from the specified distribution of correlation structures and test whether the null hypothesis is rejected under each draw. We define the \emph{breakdown statistic} as the share of correlation structures under which we are unable to reject the null hypothesis. For policymakers, the breakdown statistic provides a transparent measure of how fragile a conclusion is to uncertainty about correlations. A breakdown statistic close to zero implies that the conclusion is robust to most correlation structures, whereas a value close to one indicates that the null hypothesis is unlikely to be rejected under any plausible correlation structure. We refer to this approach of assessing how easily a conclusion ``breaks down'' under alternative correlation structures as breakdown analysis.\footnote{See \cite{manski2018right,masten2020inference,diegert2022assessing,rambachan2023more,spini2021robustness} for similar approaches.} Finally, note that if a null hypothesis can be rejected even under the worst-case correlation structure derived in Section \ref{section:general-case-inference}, the breakdown statistic must equal zero: by definition, all other admissible correlation structures imply a (weakly) smaller asymptotic variance than the worst case and therefore also lead to rejection of the null.

The exact algorithm for estimating the breakdown statistic is described in Appendix Section~\ref{section:breakdown-statistic-algorithm}; we provide a sketch of the algorithm here. We aim to assess the robustness of inference on $f({\bm\beta})$ to uncertainty about the asymptotic correlation structure of the estimated causal effects $\hat{\bm\beta} \in \mathbb{R}^d$. We define the robust region $\text{RR}_f$ as the set of admissible correlation matrices under which the null hypothesis $H_0: f(\bm\beta) < k$ is rejected at level $\alpha$:
\[
\text{RR}_f = \left\{ \rho \in \mathcal{R} : f(\hat{\bm\beta}) - z_\alpha \cdot \tau(\rho) \geq k \right\},
\]
where $\tau^2(\rho)$ denotes the asymptotic variance of $f(\hat{\bm\beta})$ under the correlation matrix $\rho$, $\mathcal{R}$ is the set of all admissible correlation matrices, and $z_\alpha$ is the $1-\alpha$ quantile of the standard normal distribution. We then define the breakdown statistic as the probability that the null is not rejected under an LKJ  prior  $\pi$ on $\rho$:
\[
{\text{BR}}_f = 1 - \Pr_{\rho \sim \pi} \left[ \rho \in \text{RR}_f \right].
\]
To estimate the Breakdown Statistic, we sample $\rho^{(1)}, \dots, \rho^{(N)}$ from the specified LKJ prior distribution $\pi$. For each draw, we compute the implied standard error $\tau^{(m)}$ and determine whether the null hypothesis is rejected. The estimated Breakdown Statistic is the proportion of draws under which we are unable to reject the null hypothesis of interest:
    \[
    \widehat{\text{BR}}_f = 1 - \frac{1}{N} \sum_{m=1}^N R^{(m)}.
    \]
In Section~\ref{section:application}, we describe how the Breakdown Statistic can help a policy-maker choose from a menu of policies to fund. 

\section{Application: Marginal Value of Public Funds}\label{section:application}

We illustrate our method by conducting inference on the Marginal Value of Public Funds (MVPF), a widely used metric for evaluating the welfare consequences of government policies. We first outline the MVPF framework and explain why our approach is particularly well suited for inference in this setting, before applying the tools developed in Sections~\ref{section:inference-procedure} and \ref{section:breakdown} to construct valid confidence intervals for MVPFs across eight policies.

\citet{hendren2020unified} popularized the Marginal Value of Public Funds (MVPF) as a unified metric for evaluating the ``bang-for-the-buck'' of public spending. An MVPF of 1 means that a policy delivers one dollar of benefits to recipients for each dollar of net government cost. Formally, the MVPF is defined as the benefits provided to recipients of a policy divided by the net cost borne by the government:
\[
MVPF = \frac{Benefits}{Net\;Government\;Costs} = \frac{\Delta W}{\Delta E - \Delta C},
\]
where $\Delta W$ denotes the estimated benefits to individuals, $\Delta E$ is the government's initial expenditure on the policy, and $\Delta C$ is the estimated reduction in government costs induced by the policy's causal effects.

Four features of the MVPF framework make our proposed method particularly well suited for valid inference. First, the MVPF is a non-linear function of multiple causal effects. To illustrate, consider the MVPF of the expanded Earned Income Tax Credit (EITC) program, Paycheck Plus. \citet{miller2017expanding} estimate the causal effects of the program on several outcomes, including earnings, employment, and after-tax income. These estimates, reported in Table~\ref{tab:paycheck_plus}, form the input vector:
\begin{align*}
   \mathbf{ \widehat{\beta}} =   \begin{bmatrix} \widehat{\beta}_1 & \widehat{\beta}_2  & \widehat{\beta}_3 & \widehat{\beta}_4 & \widehat{\beta}_5 & \widehat{\beta}_6 \end{bmatrix}'=\begin{bmatrix} 0.009 & 0.025  & 654 & 33 & 645 & 192 \end{bmatrix}'.
\end{align*}
From these causal effects, the MVPF for Paycheck Plus is constructed as
\begin{align*}
    MVPF_\text{Paycheck Plus} \equiv f(\mathbf{\widehat{\beta}}) & = \frac{1399 \times (45 - \widehat{\beta}_1) + 1364 \times 
 (34.8 - \widehat{\beta}_2)}{(\widehat{\beta}_3 - \widehat{\beta}_4) + (\widehat{\beta}_5 - \widehat{\beta}_6)} \\
    &= 0.996.
\end{align*} 

Second, in most applications, the only available information are the reported causal effect estimates and their standard errors. For example, the effects of Paycheck Plus are estimated using confidential administrative tax data, and the original study does not report the correlation structure across outcomes. Thus, the information available for inference is limited to the estimates and standard errors in Table~\ref{tab:paycheck_plus}. To conduct inference in this setting, \citet{hendren2020unified} assume a correlation structure across estimates. As we illustrate in Appendix Section~\ref{section:hsk-inference}, relying on an assumed correlation structure can imply confidence intervals that are not guaranteed to have the correct coverage. Our method ensures valid inference without requiring the variance-covariance matrix to be assumed or consistently estimated.

Third, \citet{hendren2020unified} show that reallocating spending from Policy B to Policy A is welfare-improving if and only if $MVPF_A > MVPF_B$. Testing the hypothesis $H_0: MVPF_A < MVPF_B$, then, is central to the policy choice problem.\footnote{Here, we assume that the beneficiaries of both policies receive equal welfare weights.} Our method provides a test for this hypothesis that controls size under any correlation structure.

Fourth, because the MVPF reflects the shadow price of redistribution, a welfare-maximizing government should, in principle, be willing to pay to reduce statistical uncertainty in its estimated cost of redistribution \citep{hendren2020unified}. In Section~\ref{section:sign-constraints}, we show how mild, setting-specific assumptions can be used to sharpen inference on the MVPF, offering a systematic way to reduce statistical uncertainty.

We apply our inference method to the MVPF of eight government policies spanning different domains of public expenditure: three job-training programs (Job Start, Work Advance, Year Up), two cash transfers (Paycheck Plus, Alaska Universal Basic Income), a health insurance expansion (Medicare Part D), childcare spending (foster care provision), and an unemployment insurance (UI) expansion.\footnote{The MVPFs for Job Start, Work Advance, Year Up, Paycheck Plus, and Alaska Universal Basic Income are computed in \cite{hendren2020unified}. The MVPF for Medicare Part D is computed in \cite{wettstein2020retirement}. The MVPF for foster care provision is computed in \cite{baron2022there}. The MVPF for the UI expansion is computed in \cite{huang2021welfare}.} The estimated MVPFs and 95\% confidence intervals constructed by solving ~\ref{worst-case-sdp} are shown in Figure~\ref{figure:worst-case-mvpfs}. Details of each policy and its MVPF calculation are provided in Appendix Section~\ref{section:mvpf-construction}.

Several lessons emerge from Figure~\ref{figure:worst-case-mvpfs}. First, even without assumptions on the off-diagonal entries of the variance-covariance matrix, we can reject the null that the MVPF of Job Start or Year Up exceeds one under any correlation structure, implying that a dollar spent on these job-training programs delivers less than a dollar in benefits. Second, using the variance upper bound, we test $H_0: MVPF_{\text{Alaska UBI}} < MVPF_{\text{Job Start}}$.\footnote{Since these MVPFs are based on independent samples, we assume they are uncorrelated.} We reject this null hypothesis, suggesting that reallocating spending from job-training programs to universal basic income programs could be welfare-enhancing. Finally, our estimates highlight meaningful statistical uncertainty in the relative ranking of some policies. For example, while the point estimates suggest that reallocating funds from job training to UI extensions is welfare-improving, our inference exercise shows that the uncertainty in these estimates precludes such a conclusion.

The only policy for which we have access to the underlying microdata is Medicare Part D. Using this data, we can recover the full variance-covariance matrix and compute exact confidence intervals, something that is infeasible for the other policies we study. Table~\ref{tab:medicare-part-d} compares three sets of confidence intervals for the estimated MVPF of Medicare Part D: exact intervals using the estimated correlation structure, intervals assuming all causal effects are uncorrelated, and worst-case intervals from Problem~\ref{worst-case-sdp}. The exact confidence intervals rule out MVPF values below 0.80 and above 1.95, whereas the worst-case confidence intervals rule out values below 0.17 and above 2.57. These results show that our worst-case intervals remain informative even without microdata, but also highlight a key takeaway for practitioners: reporting the estimated covariance matrix across causal effects, when feasible, can substantially improve the precision of ex-post inference.

A policymaker choosing among policies may care about whether we can robustly conclude that a policy ``pays for itself,'' rather than focusing only on the statistical uncertainty surrounding the estimated returns to each policy. To answer this question, we use the Breakdown approach from Section~\ref{section:breakdown}. The Breakdown Statistic summarizes robustness by reporting the share of admissible correlation structures under which the null hypothesis $H_0:\text{MVPF} < 1$ is not rejected. We compute this statistic in Table~\ref{tab:breakdown} using a uniform prior over the space of correlation matrices.\footnote{Because the policy-relevant threshold is $MVPF > 1$, our Breakdown Analysis focuses on policies with point estimates above 1; if the point estimate of the MVPF is less than 1, the null can never be rejected and the Breakdown Statistic is mechanically equal to 1.} Comparing Medicare Part D, Foster Care Provision, and UI Extension, we find that the conclusion that the $MVPF \geq 1$ is most robust for Foster Care Provision, which has a Breakdown Statistic of 0.67. Put differently, a Breakdown Statistic of 0.67 means that the conclusion that foster-care provision pays for itself fails to hold under roughly two-thirds of admissible correlation structures. This illustrates the value of the Breakdown Statistic: it makes clear not only whether a policy appears cost-effective, but also how fragile that conclusion is to uncertainty about the correlation structure. 

\begin{figure} 
    \begin{centering}
        \includegraphics[width = 0.8\textwidth]{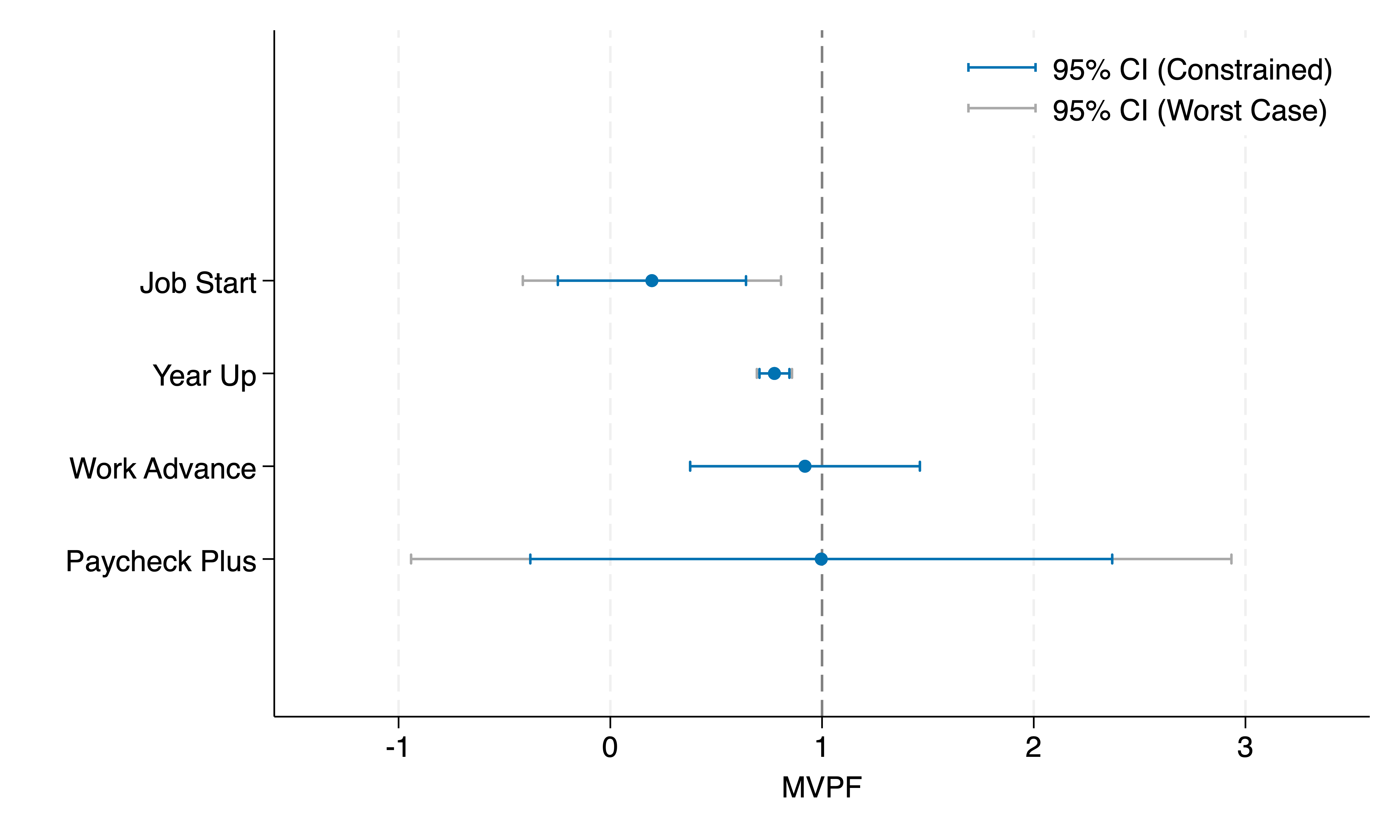}
        \caption{MVPF Confidence Intervals with Sign Constraints}\label{figure:sign-constraints-mvpf}
    \par\end{centering} 
{\scriptsize{\emph{Notes.} The figure reports the 95\% confidence intervals for the MVPF of four different policies that are evaluated using randomized trials. The confidence intervals are computed using the method described in Section \ref{section:sign-constraints}, leveraging Proposition \ref{prop:cov-result} to include sign constraints where appropriate. The construction for the MVPF of each policy as well as the sign constraints used are detailed in Appendix Section \ref{section:mvpf-construction}.}}
\end{figure}

Finally, we turn to policies evaluated using randomized trials, the setting of interest in Section~\ref{section:sign-constraints}. In these cases, Proposition~\ref{prop:cov-result} provides an interpretable characterization of the covariance structure that allows us to impose sign restrictions on correlations across outcomes. For example, in the case of Paycheck Plus, it is plausible to assume that individuals with higher after-tax income also have higher earnings and are more likely to participate in the labor force. Incorporating such restrictions, we compute the MVPF confidence intervals by solving Problem~\ref{sign-const-sdp}. Figure~\ref{figure:sign-constraints-mvpf} reports the resulting intervals for Job Start, Paycheck Plus, Work Advance, and Year Up. A key takeaway is that sign restrictions can meaningfully sharpen inference: for Paycheck Plus, the data allow us to rule out MVPF values below -0.38 and above 2.37, reducing the width of the confidence interval by nearly 30\% relative to the worst-case bound. This demonstrates how even mild, theory-motivated assumptions can deliver substantially more informative inference for policy analysis.
\clearpage
\bibliographystyle{aea}
\bibliography{references}

@article{hendren2020unified,
  title={A unified welfare analysis of government policies},
  author={Hendren, Nathaniel and Sprung-Keyser, Ben},
  journal={The Quarterly Journal of Economics},
  volume={135},
  number={3},
  pages={1209--1318},
  year={2020},
  publisher={Oxford University Press}
}

@article{cocci2023standard,
  title={Standard errors for calibrated parameters},
  author={Cocci, Matthew D and Plagborg-M{\o}ller, Mikkel},
  journal={Review of Economic Studies},
  year={2024}
}

@article{miller2017expanding,
  title={Expanding the earned income tax credit for workers without dependent children: Interim findings from the paycheck plus demonstration in new york city},
  author={Miller, Cynthia and Katz, Lawrence F and Azurdia, Gilda and Isen, Adam and Schultz, Caroline B},
  journal={New York: MDRC, September},
  year={2017}
}

@article{wettstein2020retirement,
  title={Retirement lock and prescription drug insurance: Evidence from medicare part d},
  author={Wettstein, Gal},
  journal={American Economic Journal: Economic Policy},
  volume={12},
  number={1},
  pages={389--417},
  year={2020},
  publisher={American Economic Association}
}

@article{zellner1962efficient,
  title={An efficient method of estimating seemingly unrelated regressions and tests for aggregation bias},
  author={Zellner, Arnold},
  journal={Journal of the American statistical Association},
  volume={57},
  number={298},
  pages={348--368},
  year={1962},
  publisher={Taylor \& Francis}
}

@article{masten2020inference,
  title={Inference on breakdown frontiers},
  author={Masten, Matthew A and Poirier, Alexandre},
  journal={Quantitative Economics},
  volume={11},
  number={1},
  pages={41--111},
  year={2020},
  publisher={Wiley Online Library}
}

@article{spini2021robustness,
  title={Robustness, heterogeneous treatment effects and covariate shifts},
  author={Spini, Pietro Emilio},
  journal={arXiv preprint arXiv:2112.09259},
  year={2024}
}

@article{christensen2018transparency,
  title={Transparency, reproducibility, and the credibility of economics research},
  author={Christensen, Garret and Miguel, Edward},
  journal={Journal of Economic Literature},
  volume={56},
  number={3},
  pages={920--980},
  year={2018},
  publisher={American Economic Association 2014 Broadway, Suite 305, Nashville, TN 37203-2425}
}

@article{ruggles2018historical,
  title={Historical census record linkage},
  author={Ruggles, Steven and Fitch, Catherine A and Roberts, Evan},
  journal={Annual review of sociology},
  volume={44},
  number={1},
  pages={19--37},
  year={2018},
  publisher={Annual Reviews}
}

@article{rose2018effects,
  title={The effects of job loss on crime: evidence from administrative data},
  author={Rose, Evan K},
  journal={Available at SSRN 2991317},
  year={2018}
}

@article{diegert2022assessing,
  title={Assessing omitted variable bias when the controls are endogenous},
  author={Diegert, Paul and Masten, Matthew A and Poirier, Alexandre},
  journal={arXiv preprint arXiv:2206.02303},
  year={2022}
}

@article{jones2022labor,
  title={The labor market impacts of universal and permanent cash transfers: Evidence from the Alaska Permanent Fund},
  author={Jones, Damon and Marinescu, Ioana},
  journal={American Economic Journal: Economic Policy},
  volume={14},
  number={2},
  pages={315--340},
  year={2022},
  publisher={American Economic Association 2014 Broadway, Suite 305, Nashville, TN 37203-2425}
}

@article{hendra2016encouraging,
  title={Encouraging evidence on a sector-focused advancement strategy: Two-year impacts from the WorkAdvance demonstration},
  author={Hendra, Richard and Greenberg, David H and Hamilton, Gayle and Oppenheim, Ari and Pennington, Alexandra and Schaberg, Kelsey and Tessler, Betsy L},
  journal={New York: MDRC},
  year={2016}
}

@article{fein2018bridging,
  title={Bridging the opportunity divide for low-income youth: Implementation and early impacts of the year up program},
  author={Fein, David and Hamadyk, Jill},
  journal={OPRE Report},
  volume={65},
  pages={2018},
  year={2018}
}

@article{cave1993jobstart,
  title={JOBSTART. Final Report on a Program for School Dropouts.},
  author={Cave, George and others},
  year={1993},
  publisher={ERIC},
    journal = {New York: MDRC}
}

@techreport{baron2022there,
  title={Is there a foster care-to-prison pipeline? Evidence from quasi-randomly assigned investigators},
  author={Baron, E Jason and Gross, Max},
  year={2022},
  institution={National Bureau of Economic Research}
}

@article{schaberg2017can,
  title={Can Sector Strategies Promote Longer-Term Effects? Three-Year Impacts from the WorkAdvance Demonstration.},
  author={Schaberg, Kelsey},
  journal={MDRC},
  year={2017},
  publisher={ERIC}
}

@article{huang2021welfare,
  title={The welfare effects of extending unemployment benefits: Evidence from re-employment and unemployment transfers},
  author={Huang, Po-Chun and Yang, Tzu-Ting},
  journal={Journal of Public Economics},
  volume={202},
  pages={104500},
  year={2021},
  publisher={Elsevier}
}

@misc{romano2005testing,
  title={Testing statistical hypotheses},
  author={Romano, Joseph P and Lehmann, EL},
  year={2005},
  publisher={Springer Berlin}
}

@article{elliott2015nearly,
  title={Nearly optimal tests when a nuisance parameter is present under the null hypothesis},
  author={Elliott, Graham and M{\"u}ller, Ulrich K and Watson, Mark W},
  journal={Econometrica},
  volume={83},
  number={2},
  pages={771--811},
  year={2015},
  publisher={Wiley Online Library}
}

@article{manski2018right,
  title={How do right-to-carry laws affect crime rates? Coping with ambiguity using bounded-variation assumptions},
  author={Manski, Charles F and Pepper, John V},
  journal={Review of Economics and Statistics},
  volume={100},
  number={2},
  pages={232--244},
  year={2018},
  publisher={MIT Press}
}

@article{rambachan2023more,
  title={A more credible approach to parallel trends},
  author={Rambachan, Ashesh and Roth, Jonathan},
  journal={Review of Economic Studies},
  volume={90},
  number={5},
  pages={2555--2591},
  year={2023},
  publisher={Oxford University Press US}
}

@misc{grant2014cvx,
  title={CVX: Matlab software for disciplined convex programming, version 2.1},
  author={Grant, Michael and Boyd, Stephen},
  year={2014}
}

@inproceedings{grant2008graph,
  title={Graph implementations for nonsmooth convex programs},
  author={Grant, Michael C and Boyd, Stephen P},
  booktitle={Recent advances in learning and control},
  pages={95--110},
  year={2008},
  organization={Springer}
}

@article{heckman2010rate,
  title={The rate of return to the HighScope Perry Preschool Program},
  author={Heckman, James J and Moon, Seong Hyeok and Pinto, Rodrigo and Savelyev, Peter A and Yavitz, Adam},
  journal={Journal of public Economics},
  volume={94},
  number={1-2},
  pages={114--128},
  year={2010},
  publisher={Elsevier}
}

@article{chetty2009sufficient,
  title={Sufficient statistics for welfare analysis: A bridge between structural and reduced-form methods},
  author={Chetty, Raj},
  journal={Annu. Rev. Econ.},
  volume={1},
  number={1},
  pages={451--488},
  year={2009},
  publisher={Annual Reviews}
}

\clearpage
\begin{table}[hbt!]
\centering
\caption{MVPF Calculation for Paycheck Plus}
\label{tab:paycheck_plus}
\begin{tabular}{llcc}
\toprule
& \multicolumn{1}{c}{(1)} & \multicolumn{1}{c}{(2)} &  \multicolumn{1}{c}{(3)} \\ 
\cmidrule(lr){2-4}      
& \multicolumn{1}{c}{Year} & \multicolumn{1}{c}{Estimate}  & \multicolumn{1}{c}{SE}    \\   
\midrule
         Average Bonus  Paid&  2014&  1399&\\
         Average Bonus  Paid&  2015&  1364&\\
         Take-Up&  2014&  45.90\% &\\
        Take-Up&  2015&  34.80\%&\\
         Extensive Margin Labor Market $(\widehat{\beta}_1)$&  2014&  0.90\%
&0.65\%\\
         Extensive Margin Labor Market $(\widehat{\beta}_2)$&  2015&  2.5\%
&0.91\%\\
         Impact on After Tax Income $(\widehat{\beta}_3)$&  2014&  654&187.79\\
         Impact on Earnings $(\widehat{\beta}_4)$&  2014&  33&43.35\\
          
          Impact on After Tax Income $(\widehat{\beta}_5)$ & 2015& 645&241.15 \\
           Impact on Earnings $(\widehat{\beta}_6)$ & 2015& 192&177.71\\
& & & \\
WTP &  & 1071 &  \\
Net Government Costs &  & 1074 &  \\
& & & \\
MVPF & & 0.996& \\
\midrule
\multicolumn{4}{p{5in}}{\footnotesize{\emph{Notes:} The table reports the inputs to compute the MVPF for the Paycheck Plus program. The causal effects and their corresponding standard errors are reported in \cite{miller2017expanding}. Using these estimates as inputs, the MVPF for Paycheck Plus is computed in \cite{hendren2020unified}.}}
\end{tabular} 
\end{table}

\clearpage
\newpage

\begin{table}[hbt!]
\centering
\caption{Inference for Medicare Part D MVPF}
\label{tab:medicare-part-d}
\begin{tabular}{cccc}
\toprule
 \multicolumn{1}{c}{(1)} & \multicolumn{1}{c}{(2)} & \multicolumn{1}{c}{(3)} & \multicolumn{1}{c}{(4)} \\ 
 \cmidrule(lr){1-4}      
  \multicolumn{1}{c}{MVPF} & \multicolumn{1}{c}{Exact CI} & \multicolumn{1}{c}{Independence CI} & \multicolumn{1}{c}{Worst-Case CI} \\ 
\midrule
1.37 & [0.80, 1.95] & [0.52, 2.22] & [0.18, 2.57] \\
\midrule
\multicolumn{4}{p{4in}}{\footnotesize{\emph{Notes:} The table reports 95\% confidence intervals for the MVPF of the introduction of Medicare Part D, using causal effects reported in \cite{wettstein2020retirement}. When constructing the MVPF, we use the unconditional version of the estimated causal effects of the policy on income and labor force participation for simplicity. The exact approach through which it is computed is detailed in Appendix Section \ref{section:mvpf-construction}. Column 1 reports the point estimate for the MVPF. 
Column 2 reports the exact confidence intervals for the estimated MVPF. The exact confidence intervals are computed with the Seemingly Unrelated Regression (SUR) approach of \cite{zellner1962efficient} using the (publicly available) microdata underlying the causal effects in \cite{wettstein2020retirement}. Column 3 reports the confidence intervals under the assumption that all the causal effects are uncorrelated with each other, i.e., the off-diagonal entries of the variance-covariance matrix are equal to 0. Column 4 reports the confidence intervals computed by solving \ref{worst-case-sdp}, using the method described in Section \ref{section:general-case-inference}.}}     
\end{tabular} 
\end{table}

\clearpage
\newpage

\begin{table}[hbt!]
\centering
\caption{Breakdown Statistics for MVPF}
\label{tab:breakdown}
\begin{tabular}{lc}
\toprule
 \multicolumn{1}{c}{} & \multicolumn{1}{c}{(1)} \\ 
\cmidrule(lr){2-2}      
 \multicolumn{1}{c}{} & \multicolumn{1}{c}{Breakdown Statistic} \\   
\midrule
         Medicare Part D &  0.85 \\
         Foster Care Provision & 0.67 \\
         UI Extension &   0.95 \\
\midrule
\multicolumn{2}{p{3.5in}}{\footnotesize{\emph{Notes:} The table reports Breakdown Statistic for the MVPF of Medicare Part-D \citep{wettstein2020retirement}, Foster Care Provision \citep{baron2022there}, and extension of Unemployment Insurance \citep{huang2021welfare}, using the method described in Section \ref{section:breakdown}. The reported Breakdown Statistic is calculated with respect to the null hypothesis, $H_0: MVPF < 1$. We use a uniform prior (LKJ distribution with $\eta=1$) over the space of all correlation structures.}}
\end{tabular} 
\end{table}

\clearpage
\newpage

\begin{figure} 
    \begin{centering}
        \includegraphics[width = 0.8\textwidth]{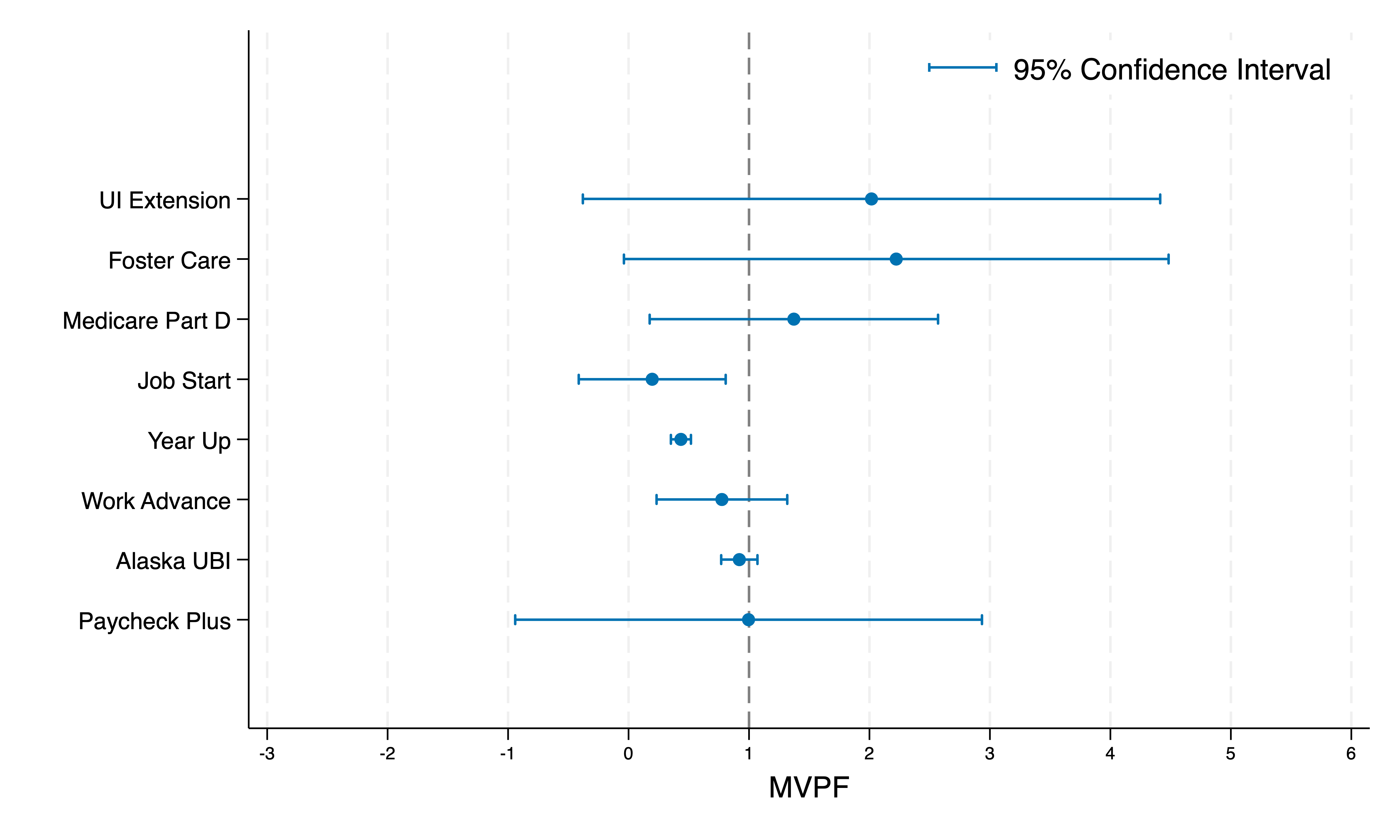}
        \caption{MVPF Confidence Intervals}\label{figure:worst-case-mvpfs}
    \par\end{centering} 
{\scriptsize{\emph{Notes.} The figure reports the 95\% confidence intervals for the MVPF of eight different policies. The construction for the MVPF of each policy is detailed in Appendix Section \ref{section:mvpf-construction}. The confidence intervals are computed using the method described in Section \ref{section:general-case-inference}, by solving \ref{worst-case-sdp}.}}
\end{figure}

\clearpage
\newpage

% \section{Conclusion}

\clearpage
\pagenumbering{arabic}% resets `page` counter to 1
\renewcommand*{\thepage}{A.\arabic{page}}
\appendix
\addcontentsline{toc}{section}{Appendix} % Add the appendix text to the document TOC
\part{Appendix} % Start the appendix part
\counterwithin{figure}{section}
\counterwithin{table}{section}
\thispagestyle{empty}
\parttoc % Insert the appendix TOC

\clearpage

\section{Connection to Existing Approaches}

\subsection{Inference Procedure in \cite{cocci2023standard}}\label{section:mikkel-inference}

In this section, we contrast our approach described in Section~\ref{section:inference-procedure} with that of \citet{cocci2023standard}, who study a related problem in the context of calibrating structural parameters to empirical moments in over-identified settings. They provide a convex optimization formulation for bounding the worst-case variance and, in Lemma~1, show that in the absence of additional restrictions, the variance can be maximized by inspecting the sign of the cross-partial term
\(
    \frac{\partial f(\bm\beta)}{\partial \beta_i}
    \frac{\partial f(\bm\beta)}{\partial \beta_j}
\).
If this product is positive, their result suggests setting $\rho_{ij}=1$ maximizes the variance, while if it is negative, the variance is maximized by setting $\rho_{ij}=-1$.

To illustrate the challenging nature of worst-case inference, even in the absence of additional constraints, consider a stylized case where inference is conducted on a function of three causal effects ($d=3$). Suppose additionally that $$
\frac{\partial f(\boldsymbol{\beta})}{\partial \beta_1} \frac{\partial f(\boldsymbol{\beta})}{\partial \beta_2}>0, \quad \frac{\partial f(\boldsymbol{\beta})}{\partial \beta_1} \frac{\partial f(\boldsymbol{\beta})}{\partial \beta_3}>0, \quad \frac{\partial f(\boldsymbol{\beta})}{\partial \beta_2} \frac{\partial f(\boldsymbol{\beta})}{\partial \beta_3}<0 .
$$

Following Lemma~1, the implied ``variance-maximizing'' correlation matrix would be:
\[
\begin{array}{c|ccc}
 & \widehat{\beta}_1 & \widehat{\beta}_2 & \widehat{\beta}_3 \\
\hline
\widehat{\beta}_1 &  1  &   1  & 1      \\
\widehat{\beta}_2 &  1  &  1  &   -1     \\
\widehat{\beta}_3 &  1  &  -1  &  1     \\

\end{array}
\]

However, this matrix has an eigenvalue equal to $-1$, meaning it is not positive semidefinite and therefore does not satisfy Constraint~\eqref{eqn:psd-const} in \ref{worst-case-sdp}. While the bound implied by such a correlation structure is a valid upper bound, a tighter upper bound can be obtained by explicitly enforcing the positive semidefiniteness constraint, as in the convex optimization problem \ref{worst-case-sdp}. This example highlights the difficulty of identifying the tightest possible worst-case variance bound, even in the absence of additional constraints.

Finally, while \citet{cocci2023standard} note that additional constraints can be incorporated directly into their convex optimization framework, it may not always be obvious to the researcher what those constraints should be. Section~\ref{section:sign-constraints} provides guidance on how to introduce such constraints in practice: for example, by exploiting structure in randomized treatment designs or by leveraging settings where treatment is random conditional on observables.

\subsection{Inference Procedure in \cite{hendren2020unified}}\label{section:hsk-inference}
To construct confidence intervals for Marginal Value of Public Funds (MVPF) estimates, \citet{hendren2020unified} adopt a parametric bootstrap procedure. They begin by specifying a correlation structure across the underlying causal effect estimates. This correlation structure is user-specified and chosen to ``maximize the width of [their] confidence intervals where estimates are from the same sample.''\footnote{The method is described in detail in Online Appendix H of \cite{hendren2020unified} as well as Section I.A of the replication files, accessible at \url{https://github.com/OpportunityInsights/welfare_analysis}.} Conditional on this specification, they repeatedly draw from a joint normal distribution centered at the reported estimates with the chosen correlation structure. For each draw, they compute the implied MVPF, generating a simulated distribution of the statistic. The 2.5th and 97.5th percentiles of this distribution are used to construct the confidence intervals. 

If the correlation structure specified in the first step of their procedure happened to coincide with the one that maximizes the width of the confidence intervals, then the bootstrap approach of \citet{hendren2020unified} would yield valid inference, and the resulting intervals would match ours. The key distinction is that our method does not assume this structure ex ante: instead, we formally identify it by solving \ref{worst-case-sdp}. Because the variance-maximizing correlation structure is rarely obvious, simply positing one does not guarantee size control. By casting the search as an optimization problem, our approach ensures valid inference regardless of the true correlation structure.

To illustrate the pitfalls of assuming a candidate worst-case correlation structure, we revisit the MVPF for Paycheck Plus, described in Section~\ref{section:application}. \citet{hendren2020unified} assume that the underlying causal effects are perfectly positively correlated, which yields confidence intervals for the MVPF of $[0.870, 1.190]$. However, solving Problem~\ref{worst-case-sdp} reveals that the true worst-case 
correlation structure is instead
\[
\begin{array}{c|cccccc}
 & \widehat{\beta}_1 & \widehat{\beta}_2 & \widehat{\beta}_3 & \widehat{\beta}_4 & \widehat{\beta}_5 & \widehat{\beta}_6 \\
\hline
\widehat{\beta}_1 &  1  &     &     &     &     &     \\
\widehat{\beta}_2 &  1  &  1  &     &     &     &     \\
\widehat{\beta}_3 &  1  &  1  &  1  &     &     &     \\
\widehat{\beta}_4 & -1  & -1  & -1  &  1  &     &     \\
\widehat{\beta}_5 &  1  &  1  &  1  & -1  &  1  &     \\
\widehat{\beta}_6 & -1  & -1  & -1  &  1  & -1  &  1  \\
\end{array}
\]
where each entry denotes the pairwise correlation $\rho_{ij} = \text{Corr}(\widehat{\beta}_i, \widehat{\beta}_j)$. Under the correct worst-case correlation shown above, the implied confidence intervals are $[-0.941, 2.934]$. This example highlights how assuming a correlation structure---even one designed to be conservative---need not deliver valid inference if it is misspecified. By contrast, our optimization-based approach guarantees size control by formally identifying the correlation structure that maximizes the variance.

A further advantage of our framework is that it avoids reliance on correlation matrices that may not be feasible. In practice, it can be difficult to tell whether a user-specified matrix respects the geometry of a valid correlation structure---specifically, whether it is positive semidefinite. As we show in Appendix Section~\ref{section:mikkel-inference}, imposing this constraint explicitly can lead to tighter confidence intervals with valid coverage rates. 

\clearpage

\section{Proposition \ref{prop:cov-result}}

\subsection{Proof of Proposition \ref{prop:cov-result}}\label{appendix:cov-proof}
We begin by writing
$$
\hat{\beta}_p=\bar{Y}_{p, 1}-\bar{Y}_{p, 0}, \quad \hat{\beta}_q=\bar{Y}_{q, 1}-\bar{Y}_{q, 0}
$$
where
$$
\bar{Y}_{p, 1}=\frac{1}{n_1} \sum_{i: Z_i=1} Y_{i p}, \quad \bar{Y}_{p, 0}=\frac{1}{n_0} \sum_{i: Z_i=0} Y_{i p}, \quad \text { and similarly for } \bar{Y}_{q, 1}, \bar{Y}_{q, 0}
$$
Then,
\begin{align*}
\Cov\left(\hat{\beta}_p, \hat{\beta}_q\right) & =\Cov\left(\bar{Y}_{p, 1}-\bar{Y}_{p, 0}, \bar{Y}_{q, 1}-\bar{Y}_{q, 0}\right) \\
& =\Cov\left(\bar{Y}_{p, 1}, \bar{Y}_{q, 1}\right)+\Cov\left(\bar{Y}_{p, 0}, \bar{Y}_{q, 0}\right)-\Cov\left(\bar{Y}_{p, 1}, \bar{Y}_{q, 0}\right)-\Cov\left(\bar{Y}_{p, 0}, \bar{Y}_{q, 1}\right)    
\end{align*}
Under random assignment and i.i.d. sampling, the treated and control groups are independent samples from the population. Therefore, 
$$
\Cov\left(\bar{Y}_{p, 1}, \bar{Y}_{q, 0}\right)=0, \quad \Cov\left(\bar{Y}_{p, 0}, \bar{Y}_{q, 1}\right)=0
$$
So:
$$
\Cov\left(\hat{\beta}_p, \hat{\beta}_q\right)=\Cov\left(\bar{Y}_{p, 1}, \bar{Y}_{q, 1}\right)+\Cov\left(\bar{Y}_{p, 0}, \bar{Y}_{q, 0}\right)
$$
We now characterize each term. Because $\left\{Y_{i p}, Y_{i q}\right\}_{i: Z_i=1}$ is an i.i.d. sample from the treated population of size $n_1$, we have:
$$
\Cov\left(\bar{Y}_{p, 1}, \bar{Y}_{q, 1}\right)=\frac{1}{n_1} \Cov\left(Y_{i p}, Y_{i q} \mid Z_i=1\right)
$$
Similarly, for the control group:
$$
\Cov\left(\bar{Y}_{p, 0}, \bar{Y}_{q, 0}\right)=\frac{1}{n_0} \Cov\left(Y_{i p}, Y_{i q} \mid Z_i=0\right)
$$
So,
$$
\Cov\left(\hat{\beta}_p, \hat{\beta}_q\right)=\frac{1}{n_1} \Cov\left(Y_{i p}, Y_{i q} \mid Z_i=1\right)+\frac{1}{n_0} \Cov\left(Y_{i p}, Y_{i q} \mid Z_i=0\right)
$$
As $n \rightarrow \infty$, the Law of Large Numbers implies:
$$
\frac{n_1}{n} \xrightarrow{p} \Prob(Z_i = 1), \quad \frac{n_0}{n} \xrightarrow{p} \Prob(Z_i = 0)
$$
So,
$$
\Cov\left(\hat{\beta}_p, \hat{\beta}_q\right) = \frac{1}{n}\left(\frac{\Cov\left(Y_{i p}, Y_{i q} \mid Z_i=1\right)}{\Prob(Z_i = 1)}+\frac{\Cov\left(Y_{i p}, Y_{i q} \mid Z_i=0\right)}{\Prob(Z_i = 0)}\right) + o_p(1)
$$
Multiplying both sides by $n$, we obtain the $(p,q)$-th entry of the asymptotic covariance matrix for $\sqrt{n}(\hat{\beta}-\beta)$ as:
$$
\frac{\Cov\left(Y_{i p}, Y_{i q} \mid Z_i=1\right)}{\Prob(Z_i = 1)}+\frac{\Cov\left(Y_{i p}, Y_{i q} \mid Z_i=0\right)}{\Prob(Z_i = 0)}
$$
which proves the proposition. $\qed$

\subsection{Extension of Proposition \ref{prop:cov-result} to Unconfoundedness}\label{appendix:unconfoundedness}

In this section, we extend the result in Proposition \ref{prop:cov-result} to the setting with covariates. Specifically, we relax the assumption in Proposition 1 that $ \Big(Y_{ij}(1),Y_{ij}(0)\Big) \perp Z_i$ for all $j=1,\dots,d$, and instead assume that 
$$
\left(Y_{i j}(1), Y_{i j}(0)\right) \perp Z_i \mid \mathbf{X}_i \quad \text{for all} \quad j = 1,\ldots,d
$$
where $\mathbf{X}_i \in \mathbb{R}^k$ is a vector of observed covariates.

Let the data $\left\{\left(Y_{i j}, Z_i, \mathbf{X}_i\right)\right\}_{i=1}^n$ be i.i.d. across units. In order to consistently estimate the average treatment effect under unconfoundedness using the oft-adopted regression estimator, we make two additional assumptions. First, we assume that there is overlap, such that the probability of being treated is bounded away from 0 and 1 at each covariate value in the support:
$$
0<\mathbb{P}\left(Z_i=1 \mid \mathbf{X}_i\right)<1 \quad \text { a.s. }
$$
We also assume that the true conditional expectation function is linear in covariates. Specifically, we assume that, for each outcome $j$,
$$
\mathbb{E}\left[Y_{i j} \mid Z_i, \mathbf{X}_i\right]=\alpha_j+\tau_j Z_i+\mathbf{X}_i^{\top} \gamma_j
$$
Let $\hat{\tau}_j$ be the OLS coefficient on $Z_i$ in a regression of $Y_{i j}$ on $Z_i$ and $\mathbf{X}_i$. In this section, we characterize the asymptotic covariance between the estimated treatment effects $\hat{\tau}_p$ and $\hat{\tau}_q$.

For each outcome $\boldsymbol{j}$, consider the linear regression:
$$
Y_{i j}=\alpha_j+\tau_j Z_i+\mathbf{X}_i^{\top} \gamma_j+\varepsilon_{i j}
$$

Define $\tilde{Z}_i:=Z_i-\Pi_Z \mathbf{X}_i$, the residual from regressing $Z_i$ on $\mathbf{X}_i$ and $\tilde{Y}_{i j}:=Y_{i j}-\Pi_j \mathbf{X}_i$, the residual from regressing $Y_{i j}$ on $\mathbf{X}_i$, where $\Pi_Z$ and $\Pi_j$ are the population projections. Then by the Frisch-Waugh-Lovell theorem, the coefficient $\hat{\tau}_j$ is equal to the slope coefficient in the regression of $\tilde{Y}_{i j}$ on $\tilde{Z}_i$, i.e.,
$$
\hat{\tau}_j=\frac{\sum_{i=1}^n \tilde{Z}_i \tilde{Y}_{i j}}{\sum_{i=1}^n \tilde{Z}_i^2}
$$
Define,
$$
w_i:=\frac{\tilde{Z}_i}{\sum_{j=1}^n \tilde{Z}_j^2}
$$
Then,
$$
\hat{\tau}_j=\sum_{i=1}^n w_i \tilde{Y}_{i j}
$$
The asymptotic covariance between the average treatment effect estimators $\hat{\tau}_p$ and $\hat{\tau}_q$ is given by,
$$
\operatorname{Cov}\left(\hat{\tau}_p, \hat{\tau}_q\right)=\operatorname{Cov}\left(\sum_{i=1}^n w_i \tilde{Y}_{i p}, \sum_{j=1}^n w_j \tilde{Y}_{j q}\right)=\sum_{i=1}^n w_i^2 \cdot \operatorname{Cov}\left(\tilde{Y}_{i p}, \tilde{Y}_{i q}\right)+\sum_{i \neq j} w_i w_j \cdot \operatorname{Cov}\left(\tilde{Y}_{i p}, \tilde{Y}_{j q}\right)
$$
Under i.i.d. sampling, $\operatorname{Cov}\left(\tilde{Y}_{i p}, \tilde{Y}_{j q}\right)=0$ for $i \neq j$, so:
$$
\operatorname{Cov}\left(\hat{\tau}_p, \hat{\tau}_q\right)=\sum_{i=1}^n w_i^2 \cdot \operatorname{Cov}\left(\tilde{Y}_{i p}, \tilde{Y}_{i q}\right)
$$
By the Law of Large Numbers, $\frac{1}{n} \sum_{i=1}^n \tilde{Z}_i^2 \xrightarrow{p} \mathbb{E}\left[\tilde{Z}_i^2\right]$ and $\frac{1}{n} \sum_{i=1}^n \tilde{Z}_i^2 \cdot \operatorname{Cov}\left(\tilde{Y}_{i p}, \tilde{Y}_{i q}\right) \xrightarrow{p} \mathbb{E}\left[\tilde{Z}_i^2 \cdot \operatorname{Cov}\left(\tilde{Y}_{i p}, \tilde{Y}_{i q}\right)\right]$. Therefore, we obtain the $(p,q)$-th entry of the asymptotic covariance matrix for $\sqrt{n}(\hat{\tau}-\tau)$ as
$$
\operatorname{Cov}\left(\hat{\tau}_p, \hat{\tau}_q\right)=\frac{\mathbb{E}\left[\tilde{Z}_i^2 \cdot \operatorname{Cov}\left(\tilde{Y}_{i p}, \tilde{Y}_{i q}\right)\right]}{\left(\mathbb{E}\left[\tilde{Z}_i^2\right]\right)^2}
$$
Thus, if
$$
\operatorname{Cov}\left(Y_{i p}, Y_{i q} \mid \mathbf{X}_i, Z_i=1\right) \geq 0 \quad \text { and } \quad \operatorname{Cov}\left(Y_{i p}, Y_{i q} \mid \mathbf{X}_i, Z_i=0\right) \geq 0 \quad \text { a.s., }
$$
then it follows that,
$$
\operatorname{Cov}\left(\hat{\tau}_p, \hat{\tau}_q\right) \geq 0 
$$ \qed

\clearpage
\section{Breakdown Statistic Computation Algorithm}\label{section:breakdown-statistic-algorithm}
In this section, we describe a step-by-step procedure to operationalize the Breakdown Approach introduced in Section~\ref{section:breakdown}.

\noindent \textbf{Algorithm:}
\begin{enumerate}
    \item Fix a null hypothesis of interest:
    \begin{align*}
    H_0: f(\bm\beta) < k 
    \quad \text{against} \quad 
    H_1: f(\bm\beta) \geq k.
\end{align*}
    \item \textbf{Compute the estimate and its gradient.}  
    Calculate $f(\hat{\bm\beta})$ and the gradient $\nabla f(\hat{\bm\beta})$.
    
    \item \textbf{Draw correlation matrices.}  
    Sample $\rho^{(1)}, \dots, \rho^{(N)} \sim \pi$, where $\pi$ is a prior distribution over the space of valid correlation matrices $\mathcal{R}$.  
    We adopt the LKJ prior \citep{lewandowski2009generating}, which has density:
    \[
    \pi(\rho) \propto \det(\rho)^{\eta - 1}.
    \]
    When $\eta = 1$, the prior is uniform over $\mathcal{R}$. Larger values of $\eta$ place more mass near the identity matrix, favoring weaker correlations.
    
    \item \textbf{Test under each draw.}  
    For each draw $\rho^{(m)}$, compute the implied standard error $\tau^{(m)}$ and determine whether the null hypothesis is rejected:
    \[
    R^{(m)} = \mathbbm{1} \!\left\{ f(\hat{\bm\beta}) - z_\alpha \cdot \tau^{(m)} \geq k \right\},
    \]
    where $z_\alpha$ is the $1-\alpha$ quantile of the standard normal distribution.
    
    \item \textbf{Estimate the breakdown statistic.}  
    Compute:
    \[
    \widehat{\text{BR}}_f = 1 - \frac{1}{N} \sum_{m=1}^N R^{(m)}.
    \]
\end{enumerate}

\noindent This statistic measures the proportion of correlation structures under which the conclusion fails to hold, assuming $\rho \sim \pi$.

\clearpage

\section{Details on MVPF Construction}\label{section:mvpf-construction}
In this Section, we detail the construction of the MVPF of all policies discussion in Section \ref{section:application}. In each case, we defer further discussion of the MVPF to the paper providing the MVPF construction for a given policy.

\subsection{Paycheck Plus}
The estimates used to construct the MVPF for Paycheck Plus are drawn from \cite{miller2017expanding}. The estimates are summarized in the following Table:
\begin{table}[hbt!]
\centering
\caption{MVPF Calculation for Paycheck Plus}
\begin{tabular}{llcc}
\toprule
& \multicolumn{1}{c}{(1)} & \multicolumn{1}{c}{(2)} &  \multicolumn{1}{c}{(3)} \\ 
\cmidrule(lr){2-4}      
& \multicolumn{1}{c}{Year} & \multicolumn{1}{c}{Estimate}  & \multicolumn{1}{c}{SE}    \\   
\midrule
         Average Bonus  Paid&  2014&  1399&\\
         Average Bonus  Paid&  2015&  1364&\\
         Take-Up&  2014&  45.90\% &\\
        Take-Up&  2015&  34.80\%&\\
         Extensive Margin Labor Market $(\widehat{\beta}_1)$&  2014&  0.90\%
&0.65\%\\
         Extensive Margin Labor Market $(\widehat{\beta}_2)$&  2015&  2.5\%
&0.91\%\\
         Impact on After Tax Income $(\widehat{\beta}_3)$&  2014&  654&187.79\\
         Impact on Earnings $(\widehat{\beta}_4)$&  2014&  33&43.35\\
          
          Impact on After Tax Income $(\widehat{\beta}_5)$ & 2015& 645&241.15 \\
           Impact on Earnings $(\widehat{\beta}_6)$ & 2015& 192&177.71\\
\midrule
\end{tabular} 
\end{table}

We replicate the construction of the MVPF for Paycheck Plus from \cite{hendren2020unified}, as follows: 
\begin{align*}
    MVPF_\text{Paycheck Plus} = f(\mathbf{\widehat{\beta}}) & = \frac{1399 \times (45 - \widehat{\beta}_1) + 1364 \times 
 (34.8 - \widehat{\beta}_2)}{(\widehat{\beta}_3 - \widehat{\beta}_4) + (\widehat{\beta}_5 - \widehat{\beta}_6)} \\
    & = 0.996.
\end{align*} 

In Figure \ref{figure:sign-constraints-mvpf}, we assume that the asymptotic correlation across all causal effects is non-negative.

\subsection{Alaska UBI}
The estimates used to construct the MVPF for Alaska UBI are drawn from \cite{jones2022labor}. 
\begin{table}[hbt!]
\centering
\caption{MVPF Calculation for Alaska UBI}
\label{tab:alaka_ubi}
\begin{tabular}{llcc}
\toprule
& \multicolumn{1}{c}{(1)} & \multicolumn{1}{c}{(2)}  \\ 
\cmidrule(lr){2-3}      
& \multicolumn{1}{c}{Estimate}  & \multicolumn{1}{c}{SE}    \\   
\midrule
         Full-Time Employment Effect ($\beta_1$) & 0.001 &  0.016 \\
         Part-Time Employment Effect ($\beta_2$) & 0.018 &  0.007 \\
\midrule
\end{tabular} 
\end{table}

We replicate the construction of the MVPF for Alaska UBI from \cite{hendren2020unified}, as follows: 
\begin{align*}
    MVPF_\text{Alaska UBI} = f(\mathbf{\widehat{\beta}}) & = \frac{1000}{1000-\Big(\beta_1 \times 5567.88 \times \frac{1000}{1602}\Big) + \Big(0.2 \times 0.5 \times \beta_2\times \frac{1000}{1602}\times 80830.57 \Big)} \\
    & = 0.92.
\end{align*} 

\subsection{Work Advance}
The estimates used to construct the MVPF for Work Advance are drawn from \cite{hendra2016encouraging} and \cite{schaberg2017can}.
\begin{table}[hbt!]
\centering
\caption{MVPF Calculation for Work Advance}
\label{tab:alaka_ubi}
\begin{tabular}{llcc}
\toprule
& \multicolumn{1}{c}{(1)} & \multicolumn{1}{c}{(2)}  \\ 
\cmidrule(lr){2-3}      
& \multicolumn{1}{c}{Estimate}  & \multicolumn{1}{c}{SE}    \\   
\midrule
         Year 2 Earnings Effect ($\beta_1$) & 1945 & 692.90 \\
         Year 3 Earnings Effect ($\beta_2$) & 1865 &	664.40 \\
\midrule
\end{tabular} 
\end{table}

We replicate the construction of the MVPF for Work Advance from \cite{hendren2020unified}, as follows: 
\begin{align*}
     MVPF_\text{Work Advance} = f(\mathbf{\widehat{\beta}}) &  \frac{\frac{\beta_1\times (1-0.003)}{1.03} + \frac{\beta_2 \times (1-0.003)}{1.03^2}}{5641-940-\beta_1 \times 0.003-\beta_2 \times 0.003} \\
     & = 0.78
\end{align*}
In Figure \ref{figure:sign-constraints-mvpf}, we assume that the asymptotic correlation across both causal effects is non-negative. 

\subsection{Year Up}
The estimates used to construct the MVPF for Year Up are drawn from \cite{fein2018bridging}.
\begin{table}[hbt!]
\centering
\caption{Year Up}
\label{tab:year-up}
\begin{tabular}{llc}
\toprule
& \multicolumn{1}{c}{(1)} & \multicolumn{1}{c}{(2)}  \\ 
\cmidrule(lr){2-3}      
& \multicolumn{1}{c}{Estimate}  & \multicolumn{1}{c}{SE}    \\   
\midrule
        Year 0 Earnings ($\beta_1$) &   -5338 & 238\\
        Year 1 Earnings ($\beta_2$) &    5181 & 474\\
        Year 2 Earnings ($\beta_3$) &    7011 & 619\\
        Discount Rate   &   3\% \\
        Tax Rate & 18.6\% \\
        Per-Participant Cost &  \$28,290 \\
        Student Stipend & \$6,614 \\
\midrule
\end{tabular}    
\end{table}

We replicate the construction of the MVPF for Year Up from \cite{hendren2020unified}, as follows: 
\begin{align*}
   MVPF_\text{Year Up} = f(\mathbf{\widehat{\beta}}) &  = \frac{(1-0.186)\times(\beta_1 + \beta_2/0.03 + \beta_3/1.03^2) + 6614}{28290-0.186\times(\beta_1 + \beta_2 + \beta_3)} \\
   & = 0.43
\end{align*}

In Figure \ref{figure:sign-constraints-mvpf}, we assume that the asymptotic correlation across $\beta_2$ and $\beta_3$ is non-negative, and the correlation of $\beta_1$ with both $\beta_2$ and $\beta_3$ is non-positive. 

\subsection{Job Start}
The estimates used to construct the MVPF for Job Start are drawn from \cite{cave1993jobstart}.

\begin{table}[hbt!]
\centering
\caption{MVPF Calculation for Job Start}
\label{tab:alaka_ubi}
\begin{tabular}{llcc}
\toprule
& \multicolumn{1}{c}{(1)} & \multicolumn{1}{c}{(2)}  \\ 
\cmidrule(lr){2-3}      
& \multicolumn{1}{c}{Estimate}  & \multicolumn{1}{c}{SE}    \\   
\midrule
Year 1 Earnings Effect ($\beta_1$) & 	-499&	151.65 \\
Year 2 Earnings Effect ($\beta_2$) &	-121&	209.20 \\
Year 3 Earnings Effect ($\beta_3$)& 	423	&258.67 \\
Year 4 Earnings Effect ($\beta_4$)&	410&	267.25 \\
Year 1 AFDC Effect ($\beta_5$)&	63	&53.96 \\
Year 2 AFDC Effect ($\beta_6$)&24	&62.94 \\
Year 3 AFDC Effect ($\beta_7$)&-3	&85.47 \\
Year 4 AFDC Effect ($\beta_8$)&-11	&84.97\\
Year 1 Food Stamps Effect ($\beta_9$)	&-45&	35.66 \\
Year 2 Food Stamps Effect ($\beta_{10}$)	&-42&	34.83 \\
Year 3 Food Stamps Effect ($\beta_{11}$)	&31	&40.94 \\
Year 4 Food Stamps Effect ($\beta_{12}$)	&31	&45.21 \\
Year 1 General Assistance Effect ($\beta_{13}$)	&24&	23.54 \\
Year 2 General Assistance Effect ($\beta_{14}$)	&7&	15.14 \\
Year 3 General Assistance Effect ($\beta_{15}$)	&-6&	24.82 \\
Year 4 General Assistance Effect ($\beta_{16}$)	&3&	26.53 \\
\midrule
\end{tabular} 
\end{table}

We replicate the construction of the MVPF for Job Start from \cite{hendren2020unified}, as follows: 
\begin{align*}
   MVPF_\text{Job Start} = f(\mathbf{\widehat{\beta}}) &  = \frac{\sum_{i=1}^{4}\beta_i\times 0.993 + \sum_{i=5}^{16}\beta_i + 606.13}{4548} \\
   & = 0.20
\end{align*}

In Figure \ref{figure:sign-constraints-mvpf}, we assume that the earnings effect in each year is negatively correlated with the AFDC, Food Stamps, and General Assistance effects in that year.

\subsection{Medicare Part D}
The estimates used to construct the MVPF for Medicare Part D are drawn from \cite{wettstein2020retirement}. The effect on labor force participation is estimated using the same specification as in Column 1, Table 1 of \cite{wettstein2020retirement}. The effect on income is estimated using the same specification as in in Column 1, Table 3 of \cite{wettstein2020retirement}. The semi-elasticity of demand for insurance is estimated using the procedure described in Appendix Section D of \cite{wettstein2020retirement}. 

\begin{table}[hbt!]
\centering
\caption{MVPF Calculation for Introduction of Medicare Part D}
\label{tab:alaka_ubi}
\begin{tabular}{llcc}
\toprule
& \multicolumn{1}{c}{(1)} & \multicolumn{1}{c}{(2)}  \\ 
\cmidrule(lr){2-3}      
& \multicolumn{1}{c}{Estimate}  & \multicolumn{1}{c}{SE}    \\   
\midrule
      Effect on Labor Force Participation ($\beta_1$) & -0.10 &  0.03 \\
      Effect on Income ($\beta_2$) & -6665.40 &  1986.92 \\
      Semi-Elasticity of Demand for Insurance ($\beta_3$) & 0.14 &  0.03 \\
\midrule
\end{tabular} 
\end{table}

We replicate the construction of the MVPF for Introduction of Medicare Part D from \cite{wettstein2020retirement}, as follows: 
\begin{align*}
   MVPF_\text{Medicare Part D} = f(\mathbf{\widehat{\beta}}) &  = \frac{0.65 \times \frac{\beta_1\times -100}{25000} \times \frac{6126}{0.4}}{(0.65+0.65\times \frac{\beta_3}{0.887} - \beta_1 - 0.28
   \times \frac{\beta_2}{1588})/0.65} \\
   & = 1.37
\end{align*}
The reason our MVPF estimate of the introduction of Medicare Part D departs from the one in \cite{wettstein2020retirement} is that we use their unconditional estimates on labor force participation and income for simplicity. 

\subsection{Foster Care}
The estimates used to construct the MVPF for Foster Care are drawn from Column 3, Table 8 in \cite{baron2022there}.
\begin{table}[hbt!]
\centering
\caption{MVPF Calculation for Foster Care}
\label{tab:alaka_ubi}
\begin{tabular}{llcc}
\toprule
& \multicolumn{1}{c}{(1)} & \multicolumn{1}{c}{(2)}  \\ 
\cmidrule(lr){2-3}      
& \multicolumn{1}{c}{Estimate}  & \multicolumn{1}{c}{SE}    \\   
\midrule
      Society's Willingness to Pay ($\beta_1$) & 83854 & 29715 \\
      Cost Savings to the Government ($\beta_2$) & 12188 &  6212 \\
\midrule
\end{tabular} 
\end{table}
We replicate the construction of the MVPF for Foster Care from \cite{baron2022there}, as follows: 
\begin{align*}
   MVPF_\text{Foster Care} = f(\mathbf{\widehat{\beta}}) &  =  \frac{\beta_1}{49920-\beta_2} \\
   & = 2.22
\end{align*}

\subsection{UI Extension}
The estimates used to construct the MVPF for UI Extension are drawn from \cite{huang2021welfare}.
\begin{table}[hbt!]
\centering
\caption{MVPF Calculation for UI Extension}
\label{tab:alaka_ubi}
\begin{tabular}{llcc}
\toprule
& \multicolumn{1}{c}{(1)} & \multicolumn{1}{c}{(2)}  \\ 
\cmidrule(lr){2-3}      
& \multicolumn{1}{c}{Estimate}  & \multicolumn{1}{c}{SE}    \\   
\midrule
      Effect on Transfers from UI ($\beta_1$) & 0.038 & 0.009 \\
      Effect on Transfers from Re-employment bonus($\beta_2$) & 0.019 & 0.011 \\
      Effect on Benefit Duration ($\beta_3$) & 56.91 & 1.96  \\
      Effect on Unemployment Duration ($\beta_4$) & 36.90 & 6.90 \\
\midrule
\end{tabular} 
\end{table}

We replicate the construction of the MVPF for extension of Unemployment Insurance from \cite{huang2021welfare}, as follows: 
\begin{align*}
   MVPF_\text{UI Extension} = f(\mathbf{\widehat{\beta}}) &  =  \frac{0.77 \times \frac{\beta_1+(\beta_2/2)}{\beta_2}+0.23}{1+(1/72.9)\times(\beta_3-55.8 - 0.5\times(\beta_3-55.8)+0.12\times\beta_4)} \\
   & = 2.02
\end{align*}

\end{document}